\documentclass[journal]{IEEEtran}
\usepackage[utf8]{inputenc}
\usepackage{array}
\newcolumntype{P}[1]{>{\centering\arraybackslash}p{#1}}
\newcolumntype{M}[1]{>{\centering\arraybackslash}m{#1}}
\usepackage{amsthm}

\usepackage{xcolor}
\usepackage{color,soul}
\usepackage{mathtools,amssymb, multicol,lipsum}
\usepackage{amsmath}

\usepackage{cuted}
\usepackage[nospread]{flushend}
\usepackage[ruled,lined]{algorithm2e}
\usepackage{float}
\usepackage{url}
\usepackage{graphicx}
\usepackage{caption}
\usepackage{subcaption}
\usepackage{flushend}
\definecolor{aqua}{rgb}{0.0, 1.0, 1.0}
\definecolor{babyblue}{rgb}{0.54, 0.81, 0.94}
\usepackage{array}
\usepackage{cite}
\ifCLASSINFOpdf
\else
\fi
\begin{document}
	%
	\title{Underwater Multi-Wavelength Optical Links With Blue Targets and Constraints: Opportunities and Challenges} 
	%
	%
\author{Alain R. Ndjiongue,~\IEEEmembership{Senior Member,~IEEE,} Octavia A. Dobre,~\IEEEmembership{Fellow,~IEEE,}
Hyundong Shin,~\IEEEmembership{Fellow,~IEEE,}
Harald Haas,~\IEEEmembership{Fellow,~IEEE}
\thanks{Alain R. Ndjiongue is with the School of Electrical and Information Engineering, Faculty of Engineering and the Built Environment, University of the Witwatersrand, Johannesburg, South Africa.} 
\thanks{Octavia A. Dobre is with the Faculty of Engineering and Applied Science, Memorial University, Canada.}
\thanks{Hyundong Shin is with the Department of Electronics and Information Convergence Engineering, Kyung Hee University, South Korea.}
\thanks{Harald Haas is the Van Eck Professor of Engineering at the University of Cambridge, Trumpington Street Cambridge CB2 1PZ UK.}}
	\maketitle
\begin{abstract}
Underwater optical wireless technologies offer multiple advantages over the acoustic technology. Acoustic signals, for instance, are susceptible to noise from marine sources due to marine life and human activities. This is not the case with optical signals. However, absorption and scattering significantly attenuate optical signals. This limits the communication range and requires higher transmission power or more sensitive receivers to detect transmitted light. Therefore, it is necessary to design underwater optical systems with a higher transmission rate and reduced attenuation. To this end, we introduce a framework for designing optical signaling constellations employing multi-wavelength light sources to account for the transmission distance and achievable rate. In particular, we redefine the color-shift keying (CSK) constraint region to target blue light and adapt to marine environments. We discuss an example of 4-point underwater CSK. The corresponding analytical results demonstrate the trade-offs between the symbol error probability, achievable rate, and transmission range of the proposed scheme. 
\end{abstract}
	\begin{IEEEkeywords}
Underwater optical wireless communication (UOWC), color shift keying (CSK), multi-wavelength (MWL) optical signals, visible light communications (VLC), and UCSK constellation design.
	\end{IEEEkeywords}
	
\IEEEpeerreviewmaketitle
	
\section{Introduction}
In 2023, the world witnessed the catastrophic implosion of Titan, a submersible from the Ocean Gate. Its mission was to take tourists to the Titanic wreck in the North Atlantic Ocean, off the coast of Newfoundland in Canada. Unfortunately, the Titan imploded during its expedition. Disrupted communication between the Titan and its support vessel contributed to the disaster. An insightful lesson from this accident is that underwater communication still has to undergo substantial improvements to be effective, particularly for high-speed data transmission over long distances. Meanwhile, a large number of applications require high-speed data transmission over short distances. Apart from long-range communications like the one between the Titan and its support vessel, which can go beyond 4 km from the sea surface, many aquatic activities rely on high-data-rate transmission systems over a short distance. An example is aquatic fauna surveillance and monitoring, which may require up to 3 Gbps of data rate to transmit data collected from sensors and cameras, depending on the resolution, compression method, and scene complexity. Other examples include critical industrial, military, and civilian applications, such as monitoring marine life, pollution detection, climate monitoring, harbor protection, and underwater archaeology. Among the solutions proposed for these high-rate and short-range underwater transmissions, we distinguish underwater optical wireless communication (UOWC) technologies, which can enable the transfer of large amounts of data between two nodes \cite{7593257}. In this paper, we propose and investigate a solution based on the transmission of multiwavelength (MWL) OWC signals in the aquatic environment.

UOWC technologies present several advantages over traditional technologies. For example, UOWCs have a minimal impact on marine life. Unlike acoustic waves, which can harm marine animals, light waves do not produce noise or vibrations. This makes it a more environmentally friendly option for underwater wireless communications. Based on these advantages and due to their potential to revolutionize aquatic communications, UOWCs have gained significant attention in recent years \cite{10704969, 10705062, 10355069}. Real-time visible light mobile communication has been demonstrated underwater with impressive results of 2 Mbps on 60 m \cite{10704969}. UOWC also has promising capabilities for shallow and turbid waters. An example is the real-time sea trial system that has been demonstrated in \cite{10705062}, promising up to 6.25 Mbps of data rate under challenging conditions such as wave fluctuation and high turbidity, over 80 m. Another example is the implementation of a 30-day UOWC system for deep-sea observation networks that has been presented in \cite{10355069}, achieving a bit rate of 125 Mbps over 30 m at a depth of 1650 m. The experiment in \cite{10355069} was characterized by its stability through a 30-day performance test, using green and blue lights in two-way UWOC links, with effective transmission of high-definition video. The study in \cite{10355069} also highlights the potential of UOWC to enhance deep-sea observation capabilities. Although the experiment in \cite{10704969} relies on a dynamic counting threshold-based system, it also supports real-time video transmission, maintaining a low rate with a moving receiver, using a 450 nm blue light to achieve transmission, while the system proposed in \cite{10355069} utilizes green and blue lights.

Although, the popularity of the UOWC is increasing, it still faces challenges in gaining widespread acceptance. One of these challenges is light attenuation in water \cite{9601292}. When light passes through water, it is scattered and absorbed, decreasing the signal strength. To reduce this impact, specialized modems have been developed \cite{10143416}, with various techniques such as forward error correction and adaptive equalization to compensate for signal loss. Another major challenge is designing and developing cost-effective and robust UOWC systems. In addition, deploying and maintaining these systems in harsh aquatic environments can be complicated and expensive. However, significant progress is necessary to ensure the advancement of UOWC technology. A way to achieve this is through the selection of an appropriate modulation scheme. In practice, schemes such as on-off keying (OOK), variable pulse position modulation, digital pulse interval modulation, and pulse width modulation have been exploited to prepare the signal for transmission \cite{7593257}. In traditional non-underwater OWC systems such as visible light communication (VLC), color shift keying (CSK) outperforms all these techniques in terms of transmission rate, as shown in \cite{8675774}. Furthermore, CSK presents the advantage of a nonvarying power signal envelope, which prevents color shifts caused by variations in data patterns, by uniformly allocating symbol constellations around the intensity of the light source, leading to a considerably reduced risk of human health complications caused by fluctuations in light intensity \cite{8675774}. For these reasons, this paper investigates the deployment of MWL optical signals in the aquatic environment, with an emphasis on CSK, also called underwater CSK (UCSK).

\begin{figure}
	\centering
	\includegraphics[width=0.99\linewidth]{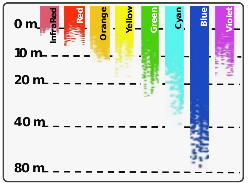}  
	\caption{Underwater MWL light propagation and attenuation \cite{8232280}}.
	\label{Fig:UnderWaterLight}
\end{figure}

Since water reacts differently to light of different colors, light color choice is fundamental in UCSK. To illustrate how lights of different colors can be affected by the underwater environment, Fig.~\ref{Fig:UnderWaterLight} depicts a MWL light propagation scenario in the underwater environment \cite{8232280}, which is obtained using a partial attenuation model. It shows how light attenuation varies with colors. Considering red-green-blue (RGB)-light emitting diodes (LEDs), Fig.~\ref{Fig:UnderWaterLight} reveals that blue light has the largest traveling distance. It travels up to 80 m, while red and green colors barely reach 5 m and 20 m, respectively \cite{8232280}. While in \cite{8232280}, all lights are transmitted with the same power, it is assumed that these transmission distances are proportional to the optical powers of the light sources, for constant underwater parameters. Therefore, the distances of 5 m, 20 m, and 80 m can be lowered or augmented by reducing or increasing the power of the corresponding LED.

Considering the impairment sources mentioned earlier that limit the optical signal transmission underwater, due to the knowledge in Fig.~\ref{Fig:UnderWaterLight} and because an analysis of UCSK has not been performed in the open literature, we present a framework to design MWL-UOWC light-based communication schemes to improve the quality of communication services in the underwater environment with a trade-off between symbol error rate (SER), achievable rate, and transmission range. The proposed MWL-UOWC system exhibits transmission rates and SER surpassing most UOWC systems.  
\begin{figure*}
	\centering
	\includegraphics[width=1.0\linewidth]{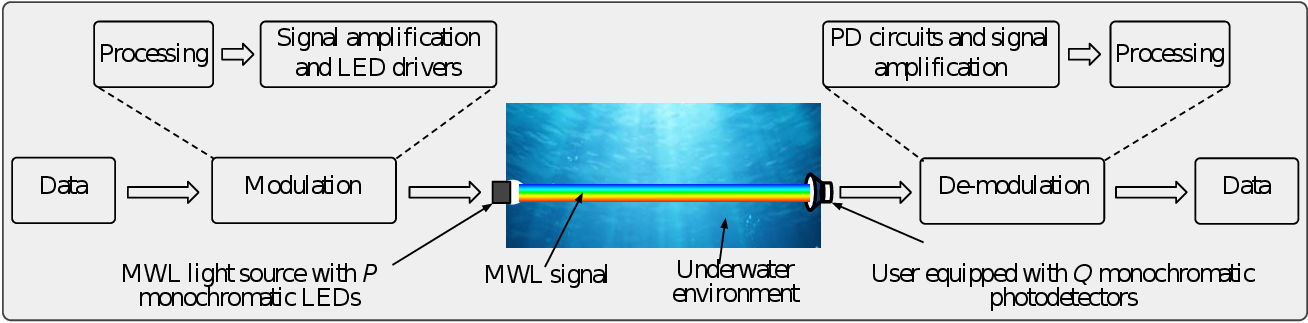} 
	\caption{A MWL-UOWC system model.}
	\label{Fig:Model}
\end{figure*}
\section{Model of the Proposed MWL-UOWC System}
\subsection{System Model}
A model of UOWC is depicted in Fig.~\ref{Fig:Model}. It shows a MWL beam crossing the underwater environment. The MWL signal is made of multiple colors lights and their combination leads to a color obtained according to the color mixing theory. At the receiver end, the colors are detected based on what is being transmitted. The data to be transmitted is treated according to the CSK technique and the resulting signals are used to control LED drivers, leading to a final color related to the combination of red, green, and blue, and their different tristimulus. This operation is done within the processing component of the transmitter, where all constraints related to the CSK scheme are integrated into the system. For example, it is in this module that the corresponding blue target is set. A general-purpose programmable processor can be used in this block, which should be capable of collecting and modulating the intensity to be converted into RGB lights. Since the intensity is related to the amount of current flowing in the LEDs, drivers are added to coordinate the output power of each color LED. These signals undergo several phenomena underwater, including attenuation and turbulence. In general, attenuation is the result of scattering and absorption. Absorption results from a thermodynamic process that depends on wavelength, while scattering is a mechanical process in which the direction of a photon is altered. Since scattering underwater is proportional to the beam angle of the light source, tight and collimated light sources will considerably reduce scattering and improve the transmission range. In practice, the white light is absorbed for more than 10\% within the first 10 m and only a little portion of the light reaches the receiver in the blue range, corroborating the underwater light propagation and attenuation demonstration in Fig.~~\ref{Fig:UnderWaterLight} \cite{8232280}. Similar to the transmitter, the receiver integrates an amplifier to increase the amount of currents induced by the detected lights to a level that can be considered by the signal processing block at the receiver. Then, the demodulation takes place to recover the transmitted data. Of course, a coding scheme can be implemented to resolve the errors on the received data \cite{8675774}.
\subsection{Class of Light Sources Used}
Solid-state electronics are advancing rapidly. This has led to the improvement of a class of LEDs, namely MWL light sources (MWL-LSs), containing several monochromatic LEDs that can be used in VLC systems \cite{6809190, mwlled}. Some off-the-shelf examples of MWL-LSs are ($a$) purple LEDs, which are a combination of a blue LED of 470 nm and a red LED of 660 nm to generate purple output light \cite{xie2022purple}; ($b$) the bi-chromatic white LED, which is amber light (598 nm) combined with cyan (505 nm); ($c$) the polychromatic LED, another type of MWL-LSs, also based on amber and cyan; ($d$) red-green LEDs, which are MWL-LSs that can provide the red only, the green only, or amber; ($e$) RGB and RGB-amber [RGB(A)] LEDs, which can generate most visible colors \cite{mwlled}. Adding amber to the RGB-LED (RGB(A)-LEDs), one obtains a quadri-chromatic MWL-LS package that outputs white light more efficiently than the conventional RGB-LEDs. RGB(A)-LEDs are used in many industries, including digital displays and lighting fixtures. The most RGB-targeted output is a white color close to daylight, which can bear the high noon sun color (R, G, B) = (1, 1, 0.984), direct sunlight color (R, G, B) = (1, 1, 1), overcast sky, (R, G, B) = (0.788, 0.886, 1), or clear blue sky (R, G, B) = (0.251, 0.612, 1). In the sequel, we focus on MWL-LSs that are capable of outputing different light colors. The design will generate colors close to blue using basic red, green, and blue LEDs, like RGB-LEDs and purple LEDs \cite{mwlled, xie2022purple}.

\begin{figure*}
     \centering
     \begin{subfigure}{0.45\textwidth}
         \centering
         \includegraphics[width=0.99\textwidth]{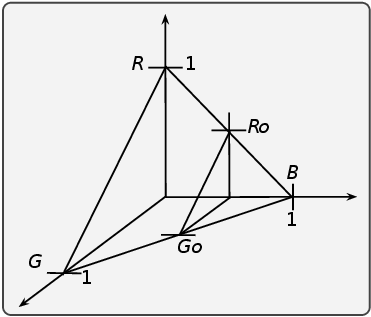}
         \caption{Constraint region for the UCSK constellation designs.}
         \label{fig:CSK_Designs}
     \end{subfigure}
     \hfill
     \begin{subfigure}{0.45\textwidth}
         \centering
         \includegraphics[width=\textwidth]{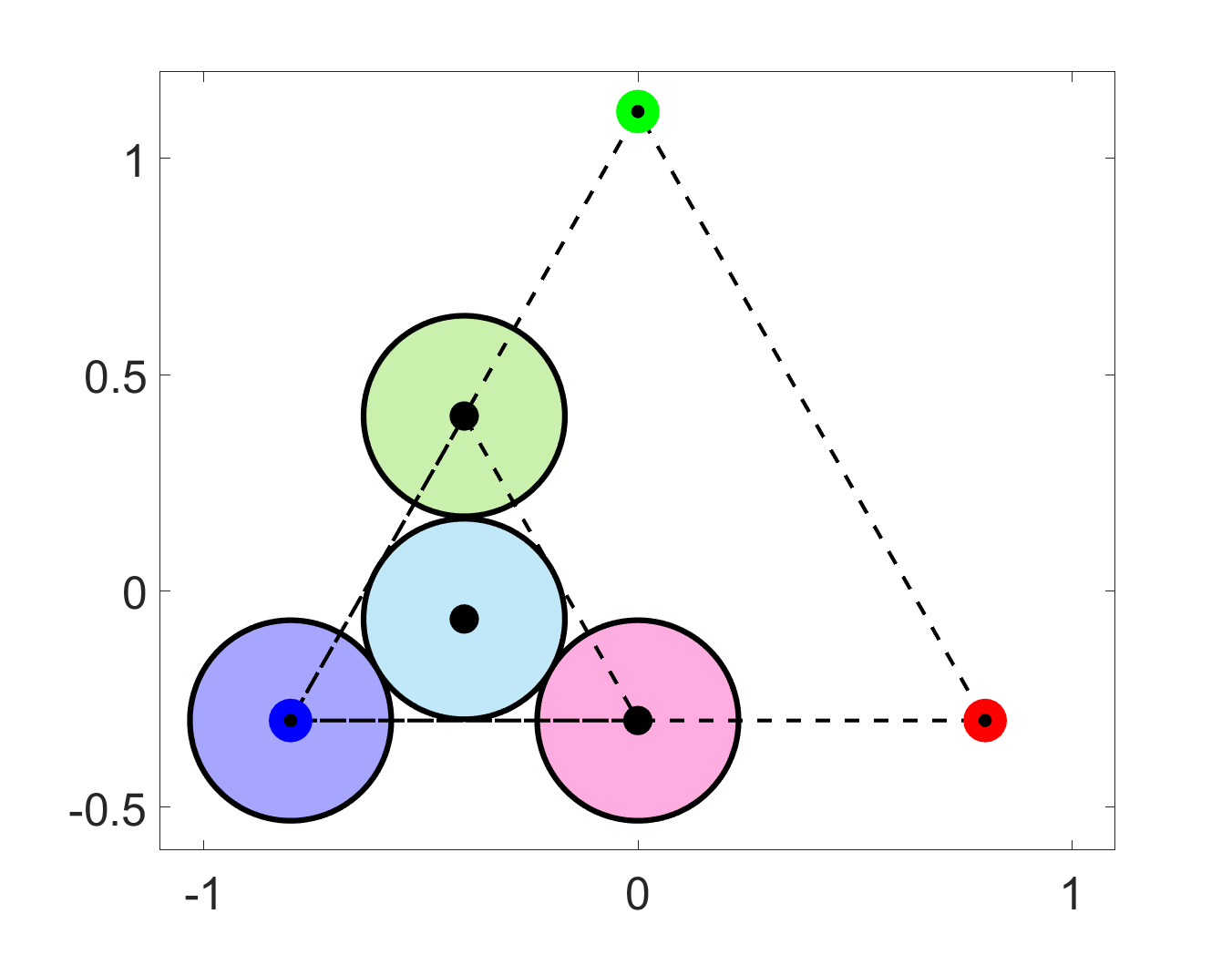}
         \caption{Generalized approach to constellation design for 4-UCSK.}
         \label{fig:4UnCSK}
     \end{subfigure}
     \caption{Constraint region for the CSK constellation designs and constellation designs for 4-UCSK.}
     \label{fig:BlueTargets}
\end{figure*}
\section{UCSK Constellation Design: Example of 4 Symbols (4-UCSK)}
As an example, we consider the design of 4-UCSK constellations, where all LEDs in the RGB package generate optimal color and power. We redefine the CSK constraint region for the underwater environment since the desired color is different from the traditional white target. This constraint region is illustrated in Fig.~\ref{fig:CSK_Designs}. The corresponding IEEE style-based symbol placement is also given in Fig.~\ref{fig:4UnCSK}. The constraint region defines the transmission conditions and helps obtaining a specific output closer to blue for an optimized transmission range. This definition keeps the blue color unchanged or fixed to its primary chromaticity. In addition, $R_o$ and $G_o$ are obtained with an optimization algorithm\footnote{In underwater VLC systems, the UCSK transmission may be classified as a mixed-integer nonlinear programming problem. For non-convex functions, it can be challenging to find multiple global optima due to the need for specialized algorithms capable of exploring the solution space thoroughly. A sequential quadratic programming (SQP) algorithm is an iterative optimization algorithm that is applied to nonlinear optimization problems with constraints like in the UCSK systems. It gradually approaches the solution by resolving a series of sub-problems in the optimization process. SQP is generally used in optimization problems in which a vector of decision variables must be minimized under constraints of equability and inequality. In our case, we consider two inequality constraints related to the output color and light intensity. Assuming that there is a narrow window of non-adverse effects of flickering on underwater fauna, we suggest that a tight variation of the transmitted light is acceptable. Consequently, we adopt the SQP solver.} based on the targeted blue color. Confining the constraint region to a smaller prism will reduce the minimum distance between symbols \cite{8697198}. However, by doing so, the system will achieve a higher transmission rate in the underwater environment while ensuring a reasonable transmission range, as opposed to other schemes like OOK. Compared to Fig.~\ref{Fig:UnderWaterLight}, when less red and green tristimulus are attributed to the red and green light sources, the final color is closer to blue. This will be less attenuated than the pure red and green colors.

\begin{figure*}
     \centering
     \begin{subfigure}{0.48\linewidth}
         \centering
         \includegraphics[width=1.2\linewidth]{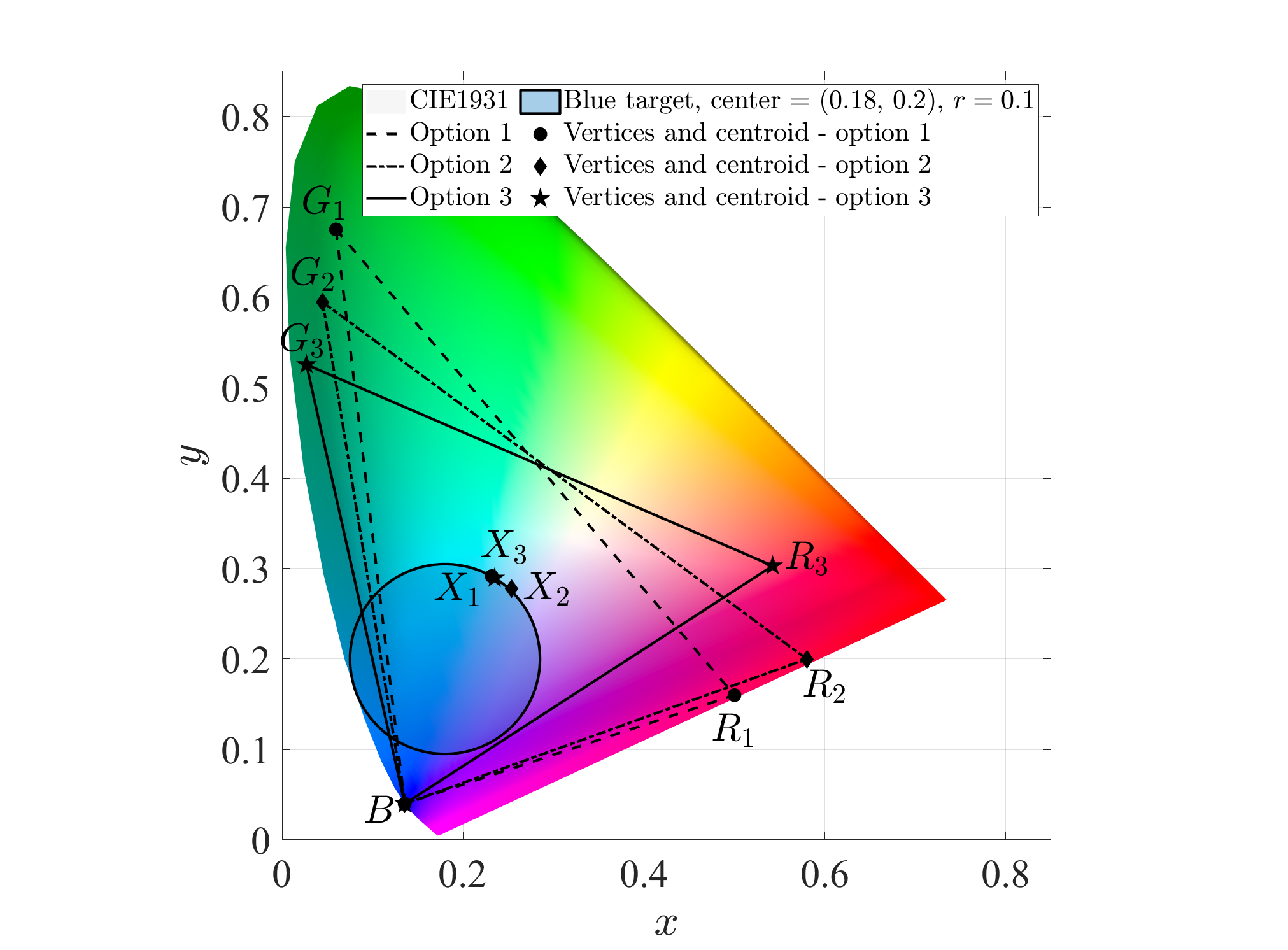}
         \caption{Blue target 1: larger spectrum, $r = 0.1$, and center = (0.15, 0.22).}
         \label{fig:Target1}
     \end{subfigure}
     \hfill
     \begin{subfigure}{0.48\linewidth}
         \centering
         \includegraphics[width=1.2\linewidth]{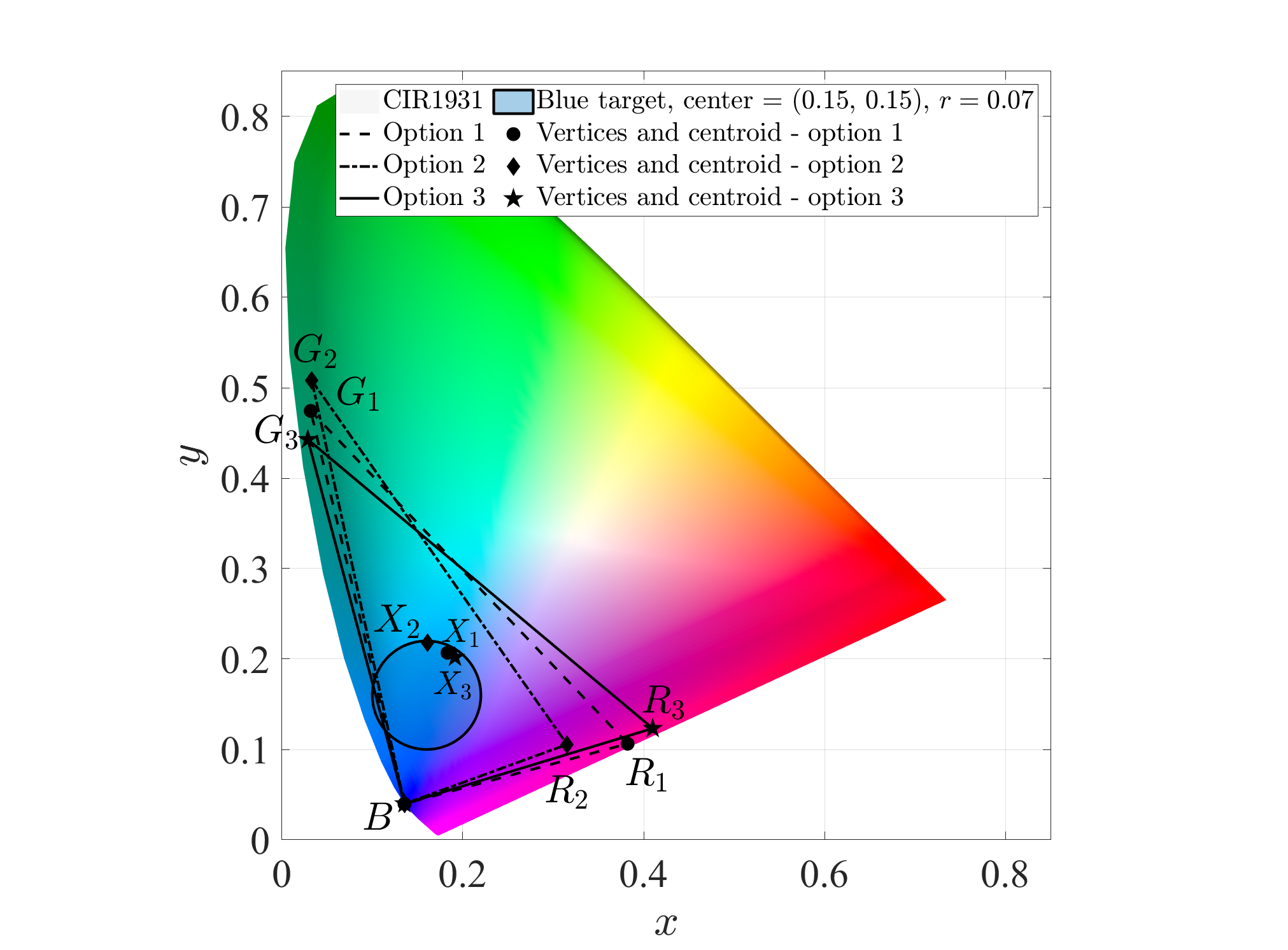}
         \caption{Blue target 2: average spectrum, $r = 0.07$, and center = (0.15, 0.15).}
         \label{fig:Target2}
     \end{subfigure}
     \hfill
     \begin{subfigure}{0.48\linewidth}
         \centering
         \includegraphics[width=1.2\linewidth]{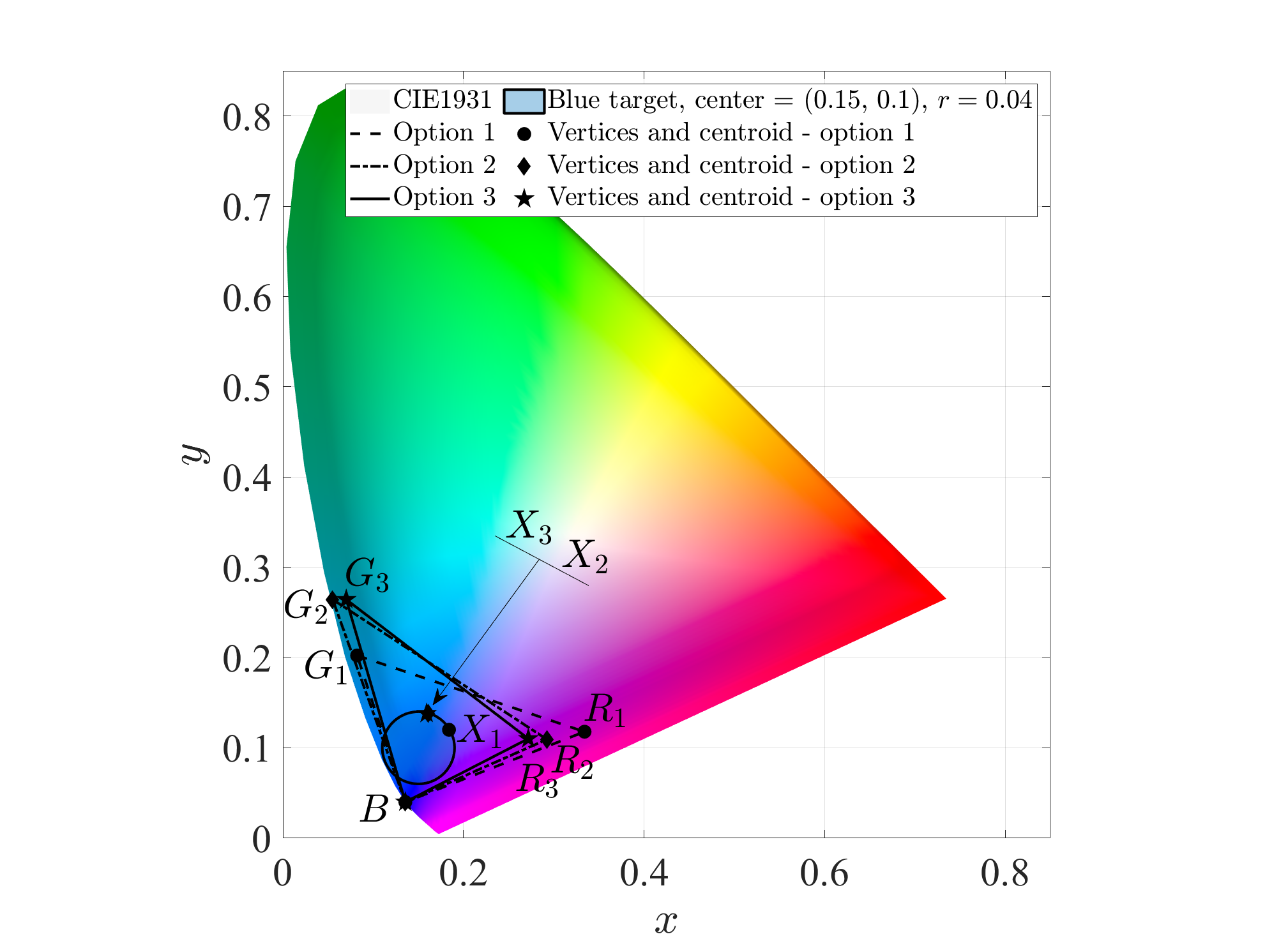}
         \caption{Blue target 3: smaller spectrum, $r = 0.04$, and center = (0.15, 0.1).}
         \label{fig:Target3}
     \end{subfigure}
    \caption{Output target color selection.\protect\footnotemark}
        \label{fig:BlueTargets}
\end{figure*}

To achieve the desired MWL transmission, we consider three blue target outputs. These target colors are selected to balance the amount of red scattered during transmission. They also improve the transmission range as the blue color is less attenuated. These three target outputs are shown on the $xy$ chromaticity diagrams depicted in Fig.~\ref{fig:BlueTargets}, where Fig.~\ref{fig:Target1}, Fig.~\ref{fig:Target2}, and Fig.~\ref{fig:Target3} respectively represent the first, second, and third blue target outputs. In Fig.~\ref{fig:Target1}, the optimization algorithm targets a circle of radius 0.1 centered at (0.18, 0.2), Fig.~\ref{fig:Target2} represents a blue in a circle of radius 0.07 located at (0.15, 0.15), and Fig.~\ref{fig:Target3} targets a blue in a circle of radius 0.04 located at (0.15, 0.1). In each of these three cases, three options are proposed, highlighting situations where the RGB LEDs in the MWL LED package may or may not generate the primary red. This will impact the detection since water will scatter all or part of the red. For example, the three options of the first blue target are materialized by: $Option$ $1$ - points $R_1$ (0.3500, 0.2000), $G_1$ (0.0594, 0.6751), $B$ (0.1355, 0.03988), $X_1$ (0.1816, 0.3050), and $d_{\text{min}} = 0.2695$; $Option$ $2$ - points $R_2$ (0.4371, 0.2709), $G_2$ (0.04482, 0.5945), $X_2$ (0.2058, 0.3018), and $d_{\text{min}} = 0.2658$; and $Option$ $3$ - points $R_3$ (0.5423, 0.3031), $G_3$ (0.02681, 0.5256), $X_3$ (0.2349, 0.2895), and $d_{\text{min}} = 0.2687$. These are shown in Table~\ref{tab12}. The same scenarios and reasoning apply to the second and third blue targets in Figs.~\ref{fig:Target2} and \ref{fig:Target3}. Note that the targeted blue colors are selected to get as close as possible to the primary blue while ensuring a combination of multiple wavelength LEDs. This explains why primary blue [$B$ (0.1355, 0.03988)] is the same in all three cases and options.

Considering $Option$ $1$ of the first blue target in Fig.~\ref{fig:Target1}, we have a lower tristimulus value for red and a higher tristimulus value for green. This scenario is characterized by the fact that the red light may be fully scattered and the corresponding receiver might not detect any signal. $Option$ $3$ is a more balanced RGB combination that provides the targeted blue. Depending on the transmission distance, the receiver will detect all three colors. $Option$ $2$ is a combination of $Options$ $1$ and $3$, giving both good color mixing and color detection.

\begin{figure*}
     \centering
     \begin{subfigure}{0.47\textwidth}
         \centering
         \includegraphics[width=\textwidth]{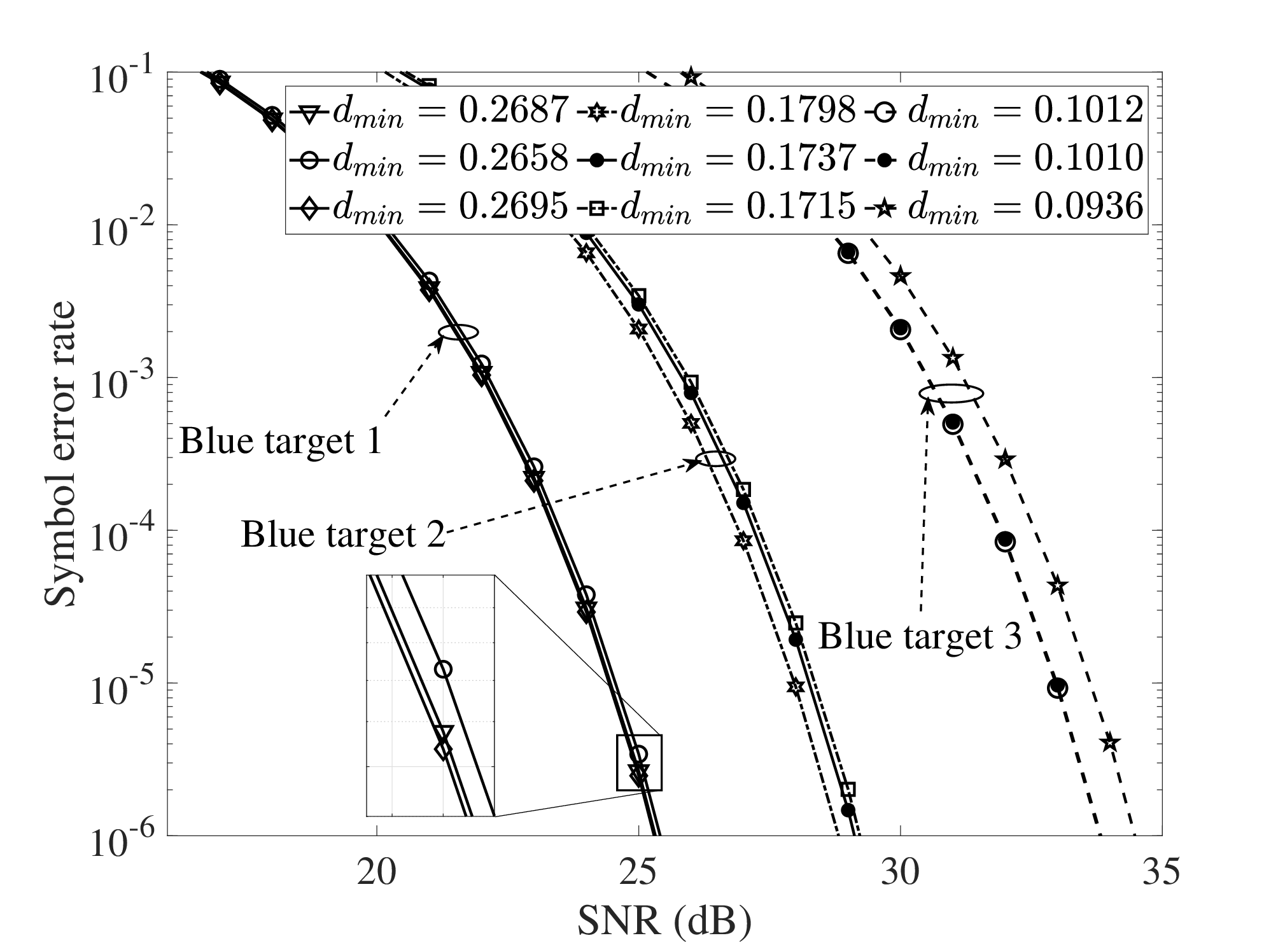}
         \caption{SER of 4-UCSK considering the three different blue targets depicted in Fig.~\ref{fig:BlueTargets}.}
         \label{fig:4UCSK}
     \end{subfigure}
     \hfill
     \begin{subfigure}{0.47\textwidth}
         \centering
         \includegraphics[width=\textwidth]{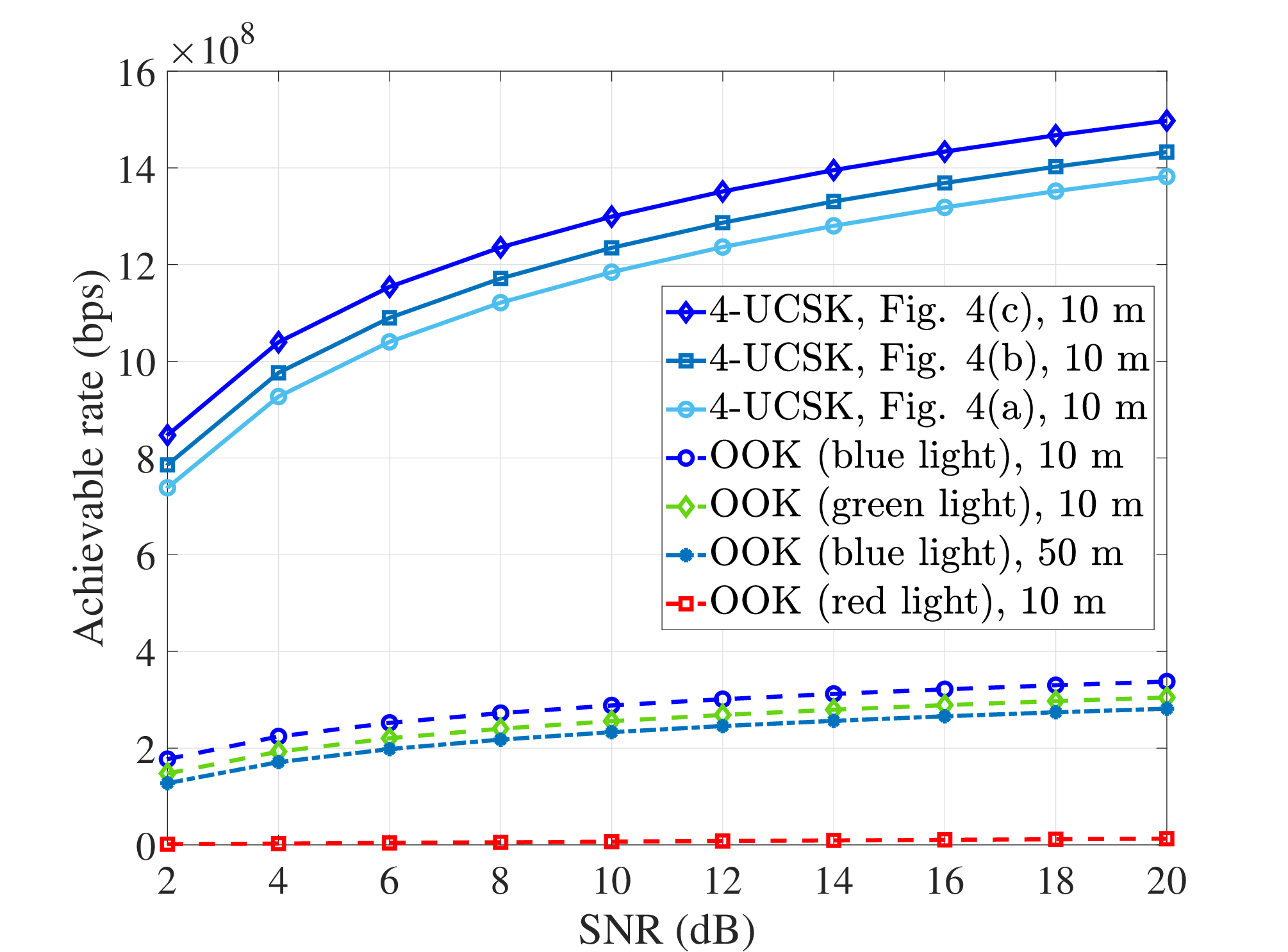}
         \caption{Achievable rates of 4-UCSK over 10 m and underwater optical OOK over 10 m and 50 m.}
         \label{fig:OOKR}
     \end{subfigure}
     \caption{SER and achievable rates of 4-UCSK considering the three different blue targets depicted in Fig.~\ref{fig:BlueTargets} over 10 m and achievable rates of an underwater optical OOK over 10 m and 50 m.}
     \label{fig:Performance}
\end{figure*}

Figure~\ref{fig:4UCSK} depicts the SER of a 4-UCSK for the three blue targets, whose design is given in Fig.~\ref{fig:BlueTargets}. It shows three proposed options for the three output targets and confirms that as we move closer to the blue color, we have less chance of detecting both green and red. However, it also shows that as the radius of the targeted blue circle is larger, the system performs better. In our design, the first blue target is the option that provides the optimal SER. As a result, the final light contains more red and green colors, which gets more scattered, leading to a shorter transmission range. The transmission distance is traded off with service quality. This contrasts with the third blue target, which exhibits longer transmission range and higher SER. The final light color is closer to blue, less prone to scattering, and leads to a higher transmission range. Here, the service quality is traded off with the transmission range. The above explanation leads to the following findings: Blue target 1 is suitable for short-range transmission systems with high communication speeds, while blue target 3 adapts well to medium- or long-range transmission systems where the SER value is not the primary requirement, and blue target 2 is the middle ground between the first and third blue targets. Blue target 2 can be used in both long- and short-range transmission systems with high- and low-service quality.

To evaluate the achievable rate of the proposed UCSK design, we first look at how the red, green, and blue light behave underwater from a practical point of view. This helps confirm our assumption about designing a UCSK with a blue target, as well as the analysis of color light propagation in Fig.~\ref{Fig:UnderWaterLight} \cite{8232280}. These achievable rates are depicted in Fig.~\ref{fig:OOKR}, where UCSK is exploited for transmission over 10 m and OOK is used for transmission over 10 m and 50 m. The system's bandwidth is set to 100 MHz. All photodetectors (PDs) are InGaAs-Positive-Intrinsic-Negative with a responsivity of 0.85 and all LEDs have an electrical-to-optical ratio of 0.55 A/W. All LEDs generate 12 lm with varying forward current and voltage combinations. The transmission environment is seawater with the absorbing and scattering characteristics given in \cite{bass2000handbook} for the entire visible light spectrum. For example, the absorption and scattering coefficients are (0.650, 0.0007), (0.0638, 0.0019), and (0.0156, 0.004) for red (700 nm), green (550 nm), and blue (460 nm), leading respectively to attenuation coefficients of 3.344 $\times$ $10^{-4}$, 4.3677 $\times$ $10^{-4}$, and 0.0101.

\footnotetext{The International Commission on Illumination (CIE) 1931 color system is a standard for color management and comparison. It is also a mathematical model that defines the relationship between the visible spectrum and uses the primary colors (red, green, and blue) to generate all colors. It measures colors' physical properties as perceived by humans.}

In Fig.~\ref{fig:OOKR}, both UCSK and OOK schemes are exploited to transmit data over 10 m and only blue OOK is used over 50 m. Over 10 m, most colors lose only 10\% of their tristimulus, allowing us to determine and compare the achievable rates of the UCSK and OOK schemes, whereas over 50 m, we can evaluate the transmission range of each color. The figure illustrates the superiority of UCSK over OOK in terms of achievable rates. It also shows that with blue target 3, we achieve higher rates than with blue targets 1 and 2. As the blue target is closer to the primary blue, the system achieves higher rates. Considering the OOK scheme, the blue light transmission offers a higher achievable rate than green and red in decreasing order. Furthermore, a transmission with blue light over 50 m has a higher rate than a transmission with red light over 10 m. Finally, the figure confirms that blue light can carry information over a longer distance than green and red lights, as also shown in \cite{8232280}.

The results for 4-UCSK, provided in Figs.~\ref{fig:4UCSK} and \ref{fig:OOKR}, confirm that the configurations in Fig.~\ref{fig:Target1} are suitable for transmission systems where a lower SER is the primary requirement. In contrast, those in Fig.~\ref{fig:Target3} tolerate higher SER, may provide a higher achievable rate, and the corresponding signal can travel a longer distance. The results related to Fig.~\ref{fig:Target2} correspond to a trade-off between SER and transmission rate. Finally, considering blue targets 1, 2, and 3, \textit{Option} 2 is a trade-off between \textit{Options} 1 and 3.

   {\renewcommand{\arraystretch}{1.4}
 \begin{table}
  \centering
    \caption{Constellation Characteristics for 4-UCSK.}
    \label{tab12}
  \begin{tabular}{M{0.5cm}|P{2.2cm}|M{2.2cm}|M{2.2cm}}
    \hline
&  \normalsize{Option 1}&\normalsize{Option 2}&\normalsize{Option 3}\\
 \hline
 \multicolumn{4}{M{7.1 cm}}{\normalsize{Blue Target 1 [$r = 0.1$, center (0.15, 0.22)]}} \\
 \hline
G & (0.0594, 0.6751) & (0.0448, 0.5945) & (0.0268, 0.5256) \\
R & (0.5000, 0.1600) & (0.5800, 0.2000) & (0.5423, 0.3031) \\
X & (0.2316, 0.2917) & (0.2058, 0.3018) & (0.2348, 0.2895) \\
$d_{\text{min}}$ & 0.2695 & 0.2658 & 0.2687 \\
 \hline
 \multicolumn{4}{M{7.1 cm}}{\normalsize{Blue Target 2 [$r = 0.06$, center (0.15, 0.15)]}} \\
 \hline
G & (0.0330, 0.5081) & (0.03185, 0.4744) & (0.02871, 0.4428) \\
R & (0.3152, 0.1055) & (0.3822, 0.1063) & (0.4099, 0.1233) \\
X & (0.1613, 0.2179) & (0.1832, 0.2069) & (0.1914, 0.2020) \\
$d_{\text{min}}$ & 0.1737 & 0.1798 & 0.1715 \\
\hline
\multicolumn{4}{M{7.1 cm}}{\normalsize{Blue Target 3 [$r = 0.04$, center (0.15, 0.1)]}} \\
 \hline
G & (0.0821, 0.2023) & (0.0550, 0.2643) & (0.0701, 0.2643) \\
R & (0.3340, 0.1178) & (0.2924, 0.1093) & (0.2717, 0.1101) \\
X & (0.1839, 0.1200) & (0.1610, 0.1379) & (0.1591, 0.1381) \\
$d_{\text{min}}$ & 0.0936 & 0.1012 & 0.1010 \\
\hline
  \end{tabular}
\end{table} } 

\section{Opportunities and Challenges}
\subsection{Opportunities}
Due to the high level of absorption, scattering, and attenuation faced by light underwater, it is evident that the solid-state industry should use many stratagems to design and develop light sources from which the output light can cross multiple layers of water over a significant distance. Since this is not an easy task, we enumerate applications that may not require a long transmission distance. In its current state, UOWC presents a number of opportunities, representing applications that do not require a long transmission range, which, however, need to manipulate large amounts of data. MWL-UOWC can be used to transmit of data collected from cameras and sensors in the underwater environment. For example, the oceanography research industry will take advantage of UCSK's high transmission rate capability to handle applications like weather forecasting, climate research, oil and gas exploitation, as well as marine geology and biology. Other useful opportunities related to MWL-UOWC are military and defense applications, the oil and gas industries, climate change monitoring, and offshore exploitations. MWL-UOWC's high-data-rate capability will further benefit underwater environmental monitoring, especially in real-time water quality and temperature evaluation for pollution level measurement.  
\subsection{Challenges}
Apart from absorption, scattering, and attenuation, the underwater environment also suffers from turbulence. This will create sliding and generate displacement at the receiver. As a matter of fact, localization of the receiver can be very difficult in sea water due to turbulence. This leads to misalignment between the transmitter and receiver. In turbid water where misalignment is unavoidable, its impact cannot be overstated. Therefore, precise localization and tracking are challenging. This challenge (misalignment) will induce error in the communication system with an impact on system SER.
\section{Future Research Directions and Conclusion}
\subsection{Multiple-Input Single-Output UCSK Systems}
Analyzing whether or not color conversion occurs underwater should be practically evaluated. As a result of color conversion (if there is color conversion), combined with the scattering phenomenon of one or more of the transmitted colors, the finally obtained or received color would have a different tristimulus. Therefore, color detection can be performed with PDs sensitive to only colors that reach the receiver. This results in a multiple-input single-output UCSK transmission system. Further investigation of this idea is necessary. Future studies should include the practical deployment of 2-, 4-UCSK, and $M$-UCSK, $M$ denoting the constellation size, for several types of water, such as sea waters, and barred ice sheets.
\subsection{Experimentation and Prototyping} 
The fact that there is a confinement of the constraint region around the blue color provides a strong motivation to implement the UCSK scheme. Therefore, at the stage where this idea stands and based on the experimental work done on UOWCs, it is evident that effectively implementing UCSK should be among the next steps. This will provide opportunities for testing and characterizing UCSK. Speed, SER, and distance tests should be performed. These practical operations on an UOWC system with the UCSK scheme represent the way forward.
\subsection{Improving Coverage}
Overall, it is critical to realize that the proposed UCSK's high transmission rate capacity is undeniable. However, UOWCs remain usable only in short range applications. UOWC's coverage should therefore continue to be expanded by pushing the technology's boundaries. Thus, we enumerate here the most significant aspects of the UOWC system that need more attention to improve the transmission range: ($a$) Two-stage injection locking techniques \cite{7778224}, used in high-power semiconductor arrays, may be applied to design special light sources for underwater VLC transmitters that achieve higher gain; ($b$) More sensitive underwater VLC receivers, such as single photon avalanched diodes, should be developed based on liquid-crystals, lensed array interfaces, photomultiplier tubes, or exploiting the single-photon avalanche technique; ($c$) The generated beam should be shaped using a small set of specific transmitter parameters to reduce or limit beam divergence and focus the light in one direction; ($d$) Line coding and/or error-correction codes can be implemented to solve errors at the receiving end when the distance becomes high and the received power is at the limit of its detection capability; ($e$) Various techniques such as narrow-band optical filters, linear equalizers, digital signal recovery modules, or pre-equalization circuits, should be explored in specific combinations to improve the quality of an UOWC signal at the receiver located at a specific distance.
\subsection{Conclusion}
This paper illustrated a framework for designing MWL optical transmission systems with blue targets underwater. The design led to the implementation of a 4-UCSK system. It considered the viscoelastic aspect of the underwater environment and the impact on light signals, with high, medium, and low scattering of red, green, and blue colors, respectively. This allowed us to define the UCSK color constraint, centered on the blue to limit scattering and consolidate the transmission range. After defining the objective function, we exploited the SQP optimization method to solve the design problem. One example was analyzed, and three options were discussed to reveal the importance of a trade-off between transmission distance and reliability. When the SER is high, which corresponds to the scenario where the MWL-LS outputs a color closer to blue, there is less scattering with an improved transmission distance. In the case of lower SERs, higher scattering occurs for the transmission signal, and the transmission distance is traded off with reliability.

\ifCLASSOPTIONcaptionsoff
\fi
\bibliographystyle{IEEEtran}
 \bibliography{Umwotoc}

\begin{thebibliography}{10}
\providecommand{\url}[1]{#1}
\csname url@samestyle\endcsname
\providecommand{\newblock}{\relax}
\providecommand{\bibinfo}[2]{#2}
\providecommand{\BIBentrySTDinterwordspacing}{\spaceskip=0pt\relax}
\providecommand{\BIBentryALTinterwordstretchfactor}{4}
\providecommand{\BIBentryALTinterwordspacing}{\spaceskip=\fontdimen2\font plus
\BIBentryALTinterwordstretchfactor\fontdimen3\font minus \fontdimen4\font\relax}
\providecommand{\BIBforeignlanguage}[2]{{%
\expandafter\ifx\csname l@#1\endcsname\relax
\typeout{** WARNING: IEEEtran.bst: No hyphenation pattern has been}%
\typeout{** loaded for the language `#1'. Using the pattern for}%
\typeout{** the default language instead.}%
\else
\language=\csname l@#1\endcsname
\fi
#2}}
\providecommand{\BIBdecl}{\relax}
\BIBdecl

\bibitem{7593257}
{Z. {Zeng} {\textit{et al.}}}, ``A survey of underwater optical wireless communications,'' \emph{IEEE Commun. Surveys \& Tuts.}, vol.~19, no.~1, pp. 204--238, First Quarter 2017.

\bibitem{10704969}
{J. {Liu} {\textit{et al.}}}, ``{Real-time underwater visible light mobile communication with an extended distance using dynamic counting threshold},'' \emph{IEEE/OSA J. Lightw. Technol.}, pp. 1--7, Oct. 2024, Early Access.

\bibitem{10705062}
{Z. {Ma} {\textit{et al.}}}, ``{Sea-trial of a real-time underwater wireless optical communication system in shallow turbid water with wave fluctuation},'' \emph{IEEE/OSA J. Lightw. Technol.}, pp. 1--10, Oct. 2024, Early Access.

\bibitem{10355069}
{J. {Zhang} {\textit{et al.}}}, ``{Long-term and real-time high-speed underwater wireless optical communications in deep sea},'' \emph{IEEE Commun. Mag.}, vol.~62, no.~3, pp. 96--101, Mar. 2024.

\bibitem{9601292}
{L. {Chen} {\textit{et al.}}}, ``{Underwater and water-air optical wireless communication},'' \emph{IEEE/OSA J. Lightw. Technol.}, vol.~40, no.~5, pp. 1440--1452, Mar. 2022.

\bibitem{10143416}
{I. {Zhilin} {\textit{et al.}}}, ``{A universal multimode (acoustic, magnetic induction, optical, RF) software defined modem architecture for underwater communication},'' \emph{IEEE Trans. Wireless Commun.}, vol.~22, no.~12, pp. 9105--9116, Dec. 2023.

\bibitem{8675774}
C.~E. Mejia and C.~N. Georghiades, ``{Coding for visible light communication using color-shift keying constellations},'' \emph{IEEE Trans. Commun.}, vol.~67, no.~7, pp. 4955--4966, Jul. 2019.

\bibitem{8232280}
T.~Łuczyński and A.~Birk, ``{Underwater image haze removal with an underwater-ready dark channel prior},'' in \emph{Proc. of OCEANS - Anchorage, AK, USA}, 18-21 Sep. 2017, pp. 1--6.

\bibitem{6809190}
{R. J. {Drost} {\textit{et al.}}}, ``{Constellation design for channel precompensation in multi-Wavelength visible light communications},'' \emph{IEEE Trans. Commun.}, vol.~62, no.~6, pp. 1995--2005, Jun. 2014.

\bibitem{mwlled}
\BIBentryALTinterwordspacing
{Marubeni America Corporation}, ``{Multi-wavelength LEDs},'' 2024, (Accessed: 14.11.2024). [Online]. Available: \url{https://tech-led.com/en/marubeni-opto/multi-wavelength-leds/.}
\BIBentrySTDinterwordspacing

\bibitem{xie2022purple}
{C. {Xie} {\textit{et al.}}}, ``{Purple light-emitting diode (LED) lights controls chlorophyll degradation and enhances nutraceutical quality of postharvest broccoli florets},'' \emph{Scientia Horticulturae}, vol. 294, pp. 1--11, Feb. 2022.

\bibitem{8697198}
``{IEEE standard for local and metropolitan area networks--Part 15.7: Short-range optical wireless communications},'' \emph{IEEE Std 802.15.7-2018 (Revision of IEEE Std 802.15.7-2011)}, pp. 1--407, Apr. 2019.

\bibitem{bass2000handbook}
M.~Bass, \emph{{Handbook of Optics}}, 1st~ed.\hskip 1em plus 0.5em minus 0.4em\relax McGraw Hill Professional, 2000, vol. III.

\bibitem{7778224}
{H. {Lu} {\textit{et al.}}}, ``{A 56 Gb/s PAM4 VCSEL-based LiFi transmission with two-stage injection-locked technique},'' \emph{IEEE Photon. J.}, vol.~9, no.~1, pp. 1--8, Feb. 2017.

\end{thebibliography}

\end{document}